\documentclass[aps,prb,10pt,floatfix,twocolumn,flushbottom]{revtex4-2} 
\usepackage{adjustbox}
\usepackage{placeins}
\usepackage[utf8]{inputenc}
\usepackage[colorlinks=true,linkcolor=magenta,urlcolor=blue,citecolor=blue]{hyperref}
\usepackage{mathtools,amsmath,amssymb,amsfonts,graphicx,tabularx,dcolumn,bm,xcolor,times,float,algorithm,algorithmic,tikz,pgf}
\allowdisplaybreaks
\usepackage{dsfont}
\usepackage{capt-of}
\usepackage{multirow}
\usepackage{CJKutf8}
\usepackage{subfig} 
\usepackage[justification=justifying]{ragged2e}

\makeatletter

\newcommand{\ttimes}{}
\newcommand{\NN}{q}
\makeatother

\DeclareMathOperator{\tr}{tr}
\usetikzlibrary{arrows.meta, decorations.markings,calc}
\usetikzlibrary{positioning}
\tikzstyle{vecArrow} = [thick, decoration={markings,mark=at position
   1 with {\arrow[semithick]{open triangle 60}}},
   double distance=1.4pt, shorten >= 5.5pt,
   preaction = {decorate},
   postaction = {draw,line width=1.4pt, white,shorten >= 4.5pt}]
\tikzstyle{innerWhite} = [semithick, white,line width=1.4pt, shorten >= 4.5pt]
\tikzset{middlearrow/.style={
        decoration={markings,
            mark= at position 0.5 with {\arrow{#1}} ,
        },
        postaction={decorate}
    }
}
\begin{document}
\begin{CJK*}{UTF8}{gbsn} 
\title{Brownian Motion in Orthogonal and Symplectic Groups}
\author{Zhiyang Tan (谭志阳)}
\author{Piet W.\ Brouwer}%
\affiliation{Dahlem Center for Complex Quantum Systems and Physics Department, Freie Universit\"at Berlin, Arnimallee 14, 14195 Berlin, Germany}
\date{\today}
\begin{abstract}
Matrix Brownian motion provides a powerful framework for studying crossover ensembles in quantum chaos and quantum transport, as well as thermalization and information scrambling in many-body dynamics. Here, we develop a unified diagrammatic framework to characterize Brownian ensembles for orthogonal and symplectic random matrices, which describe systems with particle-hole symmetry. We compute polynomial averages up to fourth order and construct an orthogonally invariant interpolation for the disconnected $\mathrm{SO}^-(q)$ sector of the orthogonal group. We consider applications relating to the fields of quantum information, quantum chaos, and quantum transport.
\end{abstract}
\maketitle
\end{CJK*}

\section{Introduction}
\label{sec:1}

Understanding how classical chaos manifests in the quantum realm
has been one of the defining themes of mathematical physics over the past
half-century. Following the Bohigas--Giannoni--Schmit (BGS)
conjecture~\cite{bohigas1984characterization}, which established that the
spectral statistics of classically chaotic systems are governed by the
Wigner--Dyson ensembles of Random Matrix Theory (RMT)~\cite{wigner1955,
dyson1962circular}, RMT emerged as the universal language of quantum chaos
\cite{guhr1998,stoeckmann1999}.
A landmark in this development was the work of Bl\"umel and Smilansky, who showed that the statistical properties of the quantum-mechanical scattering matrix of a classically chaotic system are accurately described by Dyson's circular
ensembles~\cite{blumel1990smilansky}, providing one of the first direct bridges
between these abstract ensembles and physical scattering and transport phenomena.

The dynamical mechanisms by which complex quantum systems equilibrate and approach this universal RMT regime can be elegantly described using Dyson's concept of matrix Brownian motion~\cite{dysonbrownian1962, dysonclass1972}. From the 1980s onward, this continuous-diffusion framework became a standard tool for studying transitions between different random matrix ensembles, such as crossovers between Gaussian orthogonal and unitary ensembles \cite{pandey1981statistical,Pandey_1983} and their circular counterparts \cite{Pandeyeigenvalue1991}, as well as
for analyzing
parametric level dynamics, {\em i.e.}, how spectra diffuse under continuously varying
external parameters, yielding deep insights into universal parametric
correlations and level repulsion~\cite{wilkinson1989statistical,
simons1993universal}. By interpolating between deterministic initial conditions
and invariant random-matrix ensembles in the long-time limit, matrix Brownian
motion offers a controlled route to non-equilibrium spectral dynamics.
In the context of quantum transport, Brownian motion ensembles of unitary matrices have been applied to the magnetic-field dependence of the scattering matrix of a chaotic quantum dot \cite{macedo1994,macedo1996,frahm1995c,rau1995}.
Brownian motion ensembles of unitary matrices find a natural application in the disorder-induced scattering of co-propagating chiral modes \cite{fraessdorf2016}, such as they occur in a graphene $pn$ junction in a quantizing magnetic field \cite{abanin2007}. In this case, the ``Brownian time'' $t$ represents the length $L$ of the junction.

More recently, the continuous-time framework of Brownian ensembles has been taken up well beyond spectral statistics and quantum transport of non-interacting electrons. At the interface of quantum chaos and quantum information,
phenomena such as quantum scrambling~\cite{hosur2016chaos,roberts_chaos_2017,xu2024scrambling},
operator spreading~\cite{nahum_operator_2018}, R\'enyi entanglement
entropies~\cite{nahum2017growth}, classical shadow
tomography~\cite{huang2020shadows}, and measurement-induced phase
transitions~\cite{skinner2019mipt} have become central probes of complex quantum
dynamics. While initially pioneered using discrete random unitary circuits,
continuous Brownian quantum circuits have rapidly emerged as a highly tractable
analytical counterpart for studying these same
phenomena~\cite{xu2019locality,sunderhauf2019quantum,onorati2017mixing,
jian2021syk,gerbino2024dyson,liu2024noise,tan2025operatorspreading}. This information-theoretic turn has
not bypassed the quantum-chaos community itself: Gnutzmann and Smilansky have
recently shown how information scrambling and chaos arise directly from a single
Hermitian matrix~\cite{gnutzmann2024scrambling}, connecting the
spectral-statistics program to the modern language of scrambling.

Explicitly, the Brownian motion ensemble for a $q \times q$ unitary matrix $U(t)$ is of the form
\begin{equation}
  U(t) = {\cal T}_t e^{-i \int_0^t dt' H(t')},
  \label{eq:brownian0}
\end{equation}
where $t$ is the (fictitious) time, $H(t)$ is a $q \times q$ random hermitian matrix with zero mean and delta-function Gaussian correlations, and ${\cal T}_t$ is the time-ordering prescription. 
In the long-time limit, the Brownian-motion ensemble defined by Eq.\ (\ref{eq:brownian0}) converges to the Haar-distributed circular unitary ensemble (CUE). Brownian ensembles have also been considered for symmetric and self-dual unitary matrices \cite{Pandeyeigenvalue1991,macedo1994,macedo1996,frahm1995c}, which converge to the circular orthogonal and circular symplectic ensembles (COE and CSE), respectively. These Brownian ensembles describe chaotic systems with time-reversal symmetry, for which the underlying Hamiltonian is a real or symplectic random matrix.

In this article, we consider Brownian ensembles for orthogonal or symplectic matrices $U(t)$, {\em i.e.}, unitary matrices consisting of real or quaternion numbers. These are relevant for chaotic systems with particle-hole symmetry, which have a Hamiltonian $H(t)$ that is anti-symmetric or anti-self-dual for integer or half-integer spin \cite{altland1997}, respectively, so that the corresponding scattering matrices and evolution matrices are orthogonal or symplectic. In the long-time limit, these ensembles converge to the Haar-distributed ensembles on the special orthogonal and symplectic groups, providing a dynamical route to these fundamental symmetry classes 
\cite{hunt1956,Liao_2004,dahlqvist2017}. (Haar-distributed ensembles on the full orthogonal and symplectic groups have also been referred to as ``circular real'' and ``circular quaternion ensembles '' \cite{serban2010,dahlhaus2010,beenakker2011}.) 

To characterize the Brownian ensembles, we provide an explicit expression for a polynomial average of the form 
\begin{equation}
  {\cal U}_k(t)_{i_1 \ldots i_k;i_1' \ldots i_k'} = 
  \langle U_{i_1 i_1'}(t) \ldots U_{i_k i_k'}(t) \rangle
  \label{eq:Upol}
\end{equation}
for $k=1,2,3,4$. A formal framework for the calculation of such polynomial averages for the Brownian ensembles on the orthogonal and symplectic groups was developed by Dahlqvist \cite{dahlqvist2017}, who could relate the averages for the Brownian ensembles on the orthogonal and symplectic groups to those for the unitary group. For the purpose of deriving explicit expressions, we here take a different route, generalizing the diagrammatic approach for the unitary Brownian ensemble of Refs.\ \cite{guo2024complexity, tang2024brownian, li2025graphtheoretic} to the orthogonal and symplectic cases.
While polynomial averages such as Eq.\ (\ref{eq:Upol}) do not contain the same amount of information as the full eigenvalue distribution of $U(t)$, which was calculated by Pandey and Shukla for the unitary Brownian ensemble \cite{Pandeyeigenvalue1991}, they give access to spectral form factors and are important building blocks in quantum transport applications involving superconductors \cite{dahlhaus2010,beenakker2015}, in many-body systems \cite{hosur2016chaos,roberts_chaos_2017}, in theories of operator spreading \cite{nahum_operator_2018,tan2025operatorspreading,tan2026}, and for quantities of interest in quantum information, such as the second R\'enyi entropy~\cite{hosur2016chaos,nahum2017growth} and frame potentials \cite{roberts_chaos_2017,shaya2026}.

The remainder of this article is organized as follows. 
In Sec.~\ref{sec:2} we introduce the Brownian motion ensembles for random orthogonal and symplectic matrices and we present the diagrammatic approach to calculate averages of the form of Eq.\ (\ref{eq:Upol}) as well as the trace moments.
In Sec.\ \ref{sec:4} we consider the orthogonal case in more detail. The reason is that the Brownian ensemble of Eq.~\eqref{eq:brownian0}, which has a continuous path from the identity, generates matrices of determinant $+1$ and is therefore confined to the special orthogonal group, leaving the other disconnected component of the orthogonal group~\cite{hall2003lie} --- the determinant-$(-1)$ matrices, which we denote the $\mathrm{SO}^-(q)$ sector --- inaccessible. To address this topological obstruction, in Sec.~\ref{sec:4}, we discuss application of the Brownian ensemble to a nontrivial ensemble of matrices $U(0)$ at $t=0$, which allows us to continuously interpolate between an orthogonally invariant ensemble supported on the conjugacy class of a deterministic matrix and the Haar measure. We discuss a few applications of our results in the fields of quantum information, quantum chaos, and quantum transport in Sec.\ \ref{sec:3} and conclude in Sec.\ \ref{sec:5}.


\section{Brownian ensembles for orthogonal and symplectic matrices}
\label{sec:2}

For the Brownian ensembles, one considers $q \times q$ unitary matrices $U(t)$ of the form of Eq.\ (\ref{eq:brownian0}), where $H(t)$ is chosen such that $U(t)$ is orthogonal or symplectic. 
Both cases can be treated on the same footing if we require that $U(t)$ satisfies the antiunitary involution
\begin{equation}
\label{eq:Usymm}
  U(t) = Z U(t)^{*} Z^{\dagger},
\end{equation}
where $Z Z^{\dagger} = 1$ and $Z$ is symmetric in the orthogonal case and antisymmetric in the symplectic case. The corresponding constraint for the $q \times q$ hermitian matrices $H(t)$ is
\begin{equation}
\label{eq:Hsymm}
  H(t) = -Z H(t)^{\rm T} Z^{\dagger}.
\end{equation}
The standard choices are $Z = 1$ in the orthogonal case and $Z = \sigma_2$, $\sigma_2$ being the Pauli matrix, in the symplectic case, but our expressions will be valid for arbitrary involution matrices $Z$.
For the Brownian Gaussian ensembles ~\cite{Mehta_1983,mehta1997random,dysonthreefold1962,pastur2004moments}, the matrix elements of $H(t)$ have zero average and (co)variance
\begin{align}
  \label{eq:orthvar1}
  \langle H_{ij}(t) H_{kl}(t') \rangle
  &=\,
  \left(\delta_{il}\delta_{jk} - Z_{ik} Z_{jl}^*
  \right)
  \delta(t-t'),
\end{align}
where we have chosen the unit of the Brownian time $t$ such that $H$ is dimensionless.

We will be interested in the average (\ref{eq:Upol}), which we will consider as a rank-$2k$ tensor,
\begin{equation}
  \mathcal{U}_k(t) = \langle U(t)^{\otimes k} \rangle.
\end{equation}
Following the language of the field of quantum information, we'll refer to the unitary matrix $U(t)$ as the evolution operator and to the polynomial average ${\cal U}_k(t)$ as the moment operator. Because the evolution operators in the Brownian ensemble are products of statistically uncorrelated matrices corresponding to the evolution in the infinitesimal time steps, one has $\mathcal{U}_k(t) \mathcal{U}_k(t') = \mathcal{U}_k(t+t')$, which implies that the $t$-dependence of $\mathcal{U}_k(t)$ is exponential 
\begin{equation}
  \mathcal{U}_k(t) = e^{{\cal L}_k t}.
\end{equation}
Following the calculation of the moment operator for the unitary Brownian ensembles of Refs.\ \cite{guo2024complexity, tang2024brownian, li2025graphtheoretic}, we calculate ${\cal L}_k$ by expanding $\mathcal{U}_k(t)$ for small $t$ and performing the average over the random Hamiltonian $H$. This gives \cite{dahlqvist2017}
\begin{align}
  {\cal L}_k =&\,
  - \frac{k}{2} \left(q \mp 1 \right) {\cal I}_k
  + {\cal S}_k -{\cal X}_k,
  \label{eq:Lresult}
\end{align}
where the upper and lower signs are for the orthogonal and symplectic cases, respectively, and ${\cal I}_k$, ${\cal S}_k$, and ${\cal X}_k$ are rank-$2 k$ tensors with elements
\begin{align}
  ({\cal I}_k)_{i_1,\ldots,i_k;i_1',\ldots,i_k'} =&\,
  \prod_{1 \le j \le k} \delta_{i_j i_j'}, \nonumber \\
  ({\cal S}_k)_{i_1,\ldots,i_k;i_1',\ldots,i_k'} =&\,
  \sum_{1 \le j_1 < j_2 \le k} 
  Z_{i_{j_1} i_{j_2}} Z^{*}_{i_{j_1}' i_{j_2}'}
  \prod_{j \neq j_1,j_2}^{k} \delta_{i_j i_j'}, \nonumber \\
  ({\cal X}_k)_{i_1,\ldots,i_k;i_1',\ldots,i_k'} =&\,
  \sum_{1 \le j_1 < j_2 \le k} 
  \delta_{i_{j_1} i'_{j_2}} \delta_{i_{j_1}' i_{j_2}}
  \prod_{j \neq j_1,j_2}^{k} \delta_{i_j i_j'}.
  \nonumber
\end{align}
These tensors mathematically coincide with those originally introduced by Brauer in the formulation of the Brauer algebra~\cite{braueralgebra1937}.
For $k=1$, the tensors ${\cal S}$ and ${\cal X}$ do not exist, and the second and third terms should be omitted from Eq.\ (\ref{eq:Lresult}).

To find an explicit expression for the moment operator ${\cal U}_k(t)$, it remains to calculate the exponential $e^{{\cal L}_k t}$. For $k=1$ this calculation is straightforward, because ${\cal L}_k$ is proportional to the identity, and we immediately find
\begin{equation}
  {\cal U}_1 = {\cal I}_1 e^{ -\frac{1}{2} (q \mp 1)t}.
  \label{eq:U1}
\end{equation}
Below we calculate ${\cal U}_k(t) = e^{{\cal L}_k t}$ for $k=2$, $k=3$, and $k=4$.

\subsection{The case $k= 2$}
\begin{figure}
\centering
\includegraphics[trim={ 6.7cm 18.2cm 7cm 1cm},clip,scale=0.75 ]{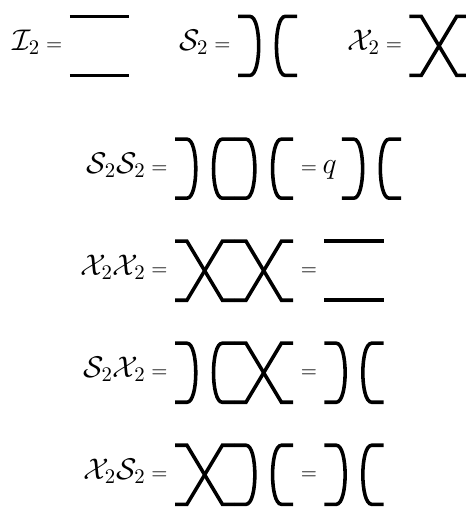}
\caption{\justifying \small \label{fig:diagrams2} Diagrammatic representation of the tensors $\mathcal{I}_2$, $\mathcal{S}_2$, and $\mathcal{X}_2$ (top) and their multiplication rules (bottom) in the orthogonal case. }
\end{figure}

\begin{figure}
\includegraphics[trim={ 6.7cm 18.2cm 7cm 1cm},clip,scale=0.75 ]{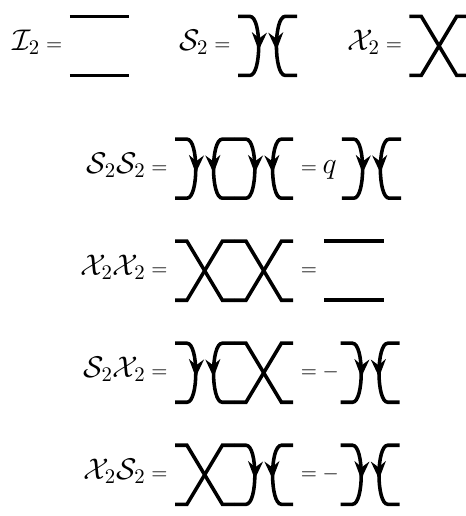}
\caption{\justifying \small \label{fig:diagramsymplectic}
Diagrammatic representation of the tensors $\mathcal{I}_2$, $\mathcal{S}_2$, and $\mathcal{X}_2$ and their multiplication rules for the symplectic (oriented) case. Note the introduction of a negative sign in the product of $\mathcal{S}_2$ and $\mathcal{X}_2$, which follows from reversing the orientation of the solid line on the left end of the diagram.}
\end{figure}
For $k=2$, the rank-$4$ tensors ${\cal I}_2$, ${\cal S}_2$, and ${\cal X}_2$ form a commutative algebra with the nontrivial multiplication rules
\begin{equation}
  \label{eq:mult2}
  {\cal S}_2^2 = q {\cal S}_2, \ \
  {\cal X}_2^2 = {\cal I}_2, \ \ 
  {\cal S}_2 {\cal X}_2 = {\cal X}_2 {\cal S}_2 = 
  \pm {\cal S}_2,
\end{equation}
where the upper sign refers to the orthogonal case and the lower sign to the symplectic case. We further made use of the identity $\mbox{tr}\, Z Z^{\dagger} = q$ and used the symmetry/antisymmetry of $Z$ in the orthogonal/symplectic cases, respectively.
The three rank-$4$ tensors ${\cal I}_2$, ${\cal S}_2$, and ${\cal X}_2$ and the multiplication rules are visualized diagrammatically in Figs.\ \ref{fig:diagrams2} and \ref{fig:diagramsymplectic}. In both figures, the Kronecker deltas are represented by solid lines connecting the left and right parts of the diagram. Elements of $Z$ are represented by solid lines connecting the left or right parts to themselves. In the symplectic case, which is shown in Fig.\ \ref{fig:diagramsymplectic}, $Z$ is antisymmetric, so that the corresponding solid lines are oriented, whereby flipping the orientation gives a minus sign. In the orthogonal case, see Fig.\ \ref{fig:diagrams2}, $Z$ is symmetric, so that no orientation is necessary. Multiplication of two tensors amounts to contraction of the adjacent endpoints of the diagrams corresponding to these tensors, whereby a closed loop contributes a factor of $q$ if its constituent lines point in the same direction ({\em e.g.}, both downward), and $-q$ if they point in opposite directions (one upward, one downward). This is how one arrives at the multiplication rules of Eq.\ (\ref{eq:mult2}).

With the help of the multiplication rules of Eq.\ (\ref{eq:mult2}), left multiplication by ${\cal L}_2$ can be represented as a matrix multiplication,
\begin{equation}
  \label{eq:M2def}
  {\cal L}_2 \begin{pmatrix} {\cal I}_2 \\ {\cal S}_2 \\ {\cal X}_2 \end{pmatrix}
  = M \begin{pmatrix} {\cal I}_2 \\ {\cal S}_2 \\ {\cal X}_2 \end{pmatrix},
\end{equation}
with
\begin{equation}
  \label{eq:M2}
  M_2 = 
\begin{pmatrix}
  \pm 1 - q  & 1 & -1 \\
  0 & 0 & 0 \\
  - 1 & \pm1 & \pm 1 - q
\end{pmatrix}.
\end{equation}
The  moment operator ${\cal U}_2(t) = e^{{\cal L}_2 t}$ can now be calculated as
\begin{align}
  e^{{\cal L}_2 t} =&\, \begin{pmatrix} 1 & 0 & 0 \end{pmatrix}
  e^{M_2 t} \begin{pmatrix} {\cal I}_2 \\ {\cal S}_2 \\ {\cal X}_2 \end{pmatrix}.
\end{align}
Using the eigenvalue-eigenvector decomposition of $M_2$ we then easily find
\begin{align}
  \label{eq:U2}
  {\cal U}_2(t) =&\,
  \frac{1}{2} e^{-(q \mp 1) t}
  \left[ 
  \vphantom{\frac{M}{M}}
  (e^{ t} + e^{- t}) {\cal I}_2
  \nonumber \right. \\ &\, \left. \mbox{}
  + \frac{2}{q} (e^{(q \mp 1) t} - e^{\mp t}) {\cal S}_2
  - (e^{ t} - e^{- t}) {\cal X}_2 \right].
\end{align}
The moment operators ${\cal U}_k(t)$ for the orthogonal and symplectic cases for $k=2$ are related by the duality transformation $t \to -t$, $q \to -q$, ${\cal S}_2 \to -{\cal S}_2$, ${\cal X}_2\to -{\cal X}_2$.

\subsection{The case $k=3$}
\begin{figure*}[tbp]
        \centering
        \includegraphics[scale=0.7,trim={3cm 16cm 4cm 1cm},clip]{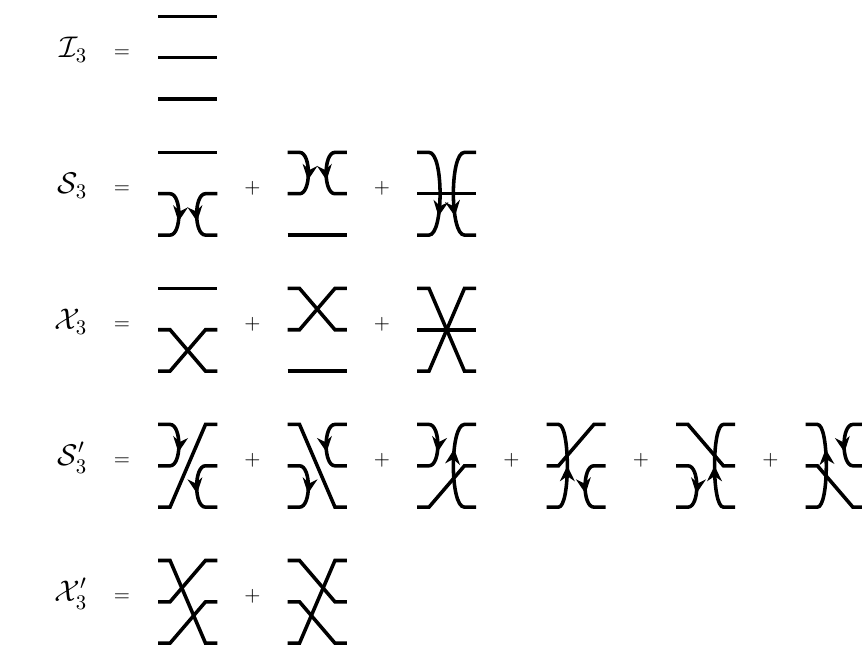}
\caption{\justifying \small Diagrammatic representations of the tensors ${\cal I}_3$, ${\cal S}_3$, ${\cal X}_3$, ${\cal S}'_3$, and ${\cal X}'_3$ in the symplectic case. Diagrammatic representations for the orthogonal case are obtained by omitting the orientations indicated by the arrows (compare Figs.\ \ref{fig:diagrams2} and \ref{fig:diagramsymplectic}).} 
    \label{fig:combined_diagrams}
\end{figure*}
\begin{table}
\begin{tabular*}{\linewidth}{c@{\extracolsep{\fill}} cccc}
 \hline\hline
$\times$ & ${\cal S}$ & ${\cal X}$ & ${\cal S}'$ & ${\cal X}'$ \\
\hline
  ${\cal S}$
  & $q {\cal S} \pm {\cal S}'$ & $\pm{\cal S} \pm{\cal S}'$ &
    $ ( q\pm1) {\cal S}'\pm2 {\cal S} $ & ${\cal S}'$ \\
  ${\cal X}$
  &  $\pm{\cal S}\pm {\cal S}'$ & $3 {\cal I} + 3 {\cal X}'$ &
    $\pm2 {\cal S} \pm 2 {\cal S}'$ & $2 {\cal X}$ \\
  ${\cal S}'$
  & $ (q\pm1) {\cal S}'\pm2 {\cal S}$ & $\pm2 {\cal S} \pm 2 {\cal S}'$ &
    $(2 q \pm 2) {\cal S} + (q \pm 3) {\cal S}'$ &
    $2 {\cal S} + {\cal S}'$ \\
  ${\cal X}'$
  & ${\cal S}'$ & $2{\cal X}$ & $2 {\cal S} + {\cal S}'$ &
    $2 {\cal I} + {\cal X}'$ \\
\hline \hline
\end{tabular*}
\caption{\justifying \small
Multiplication rules for the algebra spanned by ${\cal I}_3$, ${\cal S}_3$, and ${\cal X}_3$ (omitting the label $_k$ with $k=3$ for symplicity of notation). Diagrammatric expressions for the operators ${\cal I}_3$, ${\cal S}_3$, ${\cal X}_3$, ${\cal S}'_3$, and ${\cal X}'_3$ are shown in Fig.\ \ref{fig:combined_diagrams}.}
\label{tab:mult3}
\vspace{-0.5cm}
\end{table}

For $k=3$, products involving ${\cal S}_3$ and ${\cal X}_3$ give additional tensors that are linearly independent of ${\cal I}_3$, ${\cal S}_3$, and ${\cal X}_3$. In total, the commutative algebra generated by ${\cal I}_3$, ${\cal S}_3$, and ${\cal X}_3$ has five linearly independent basis elements. The two additional rank-$6$ tensors are denoted ${\cal X}_3'$ and ${\cal S}'_3$,
\begin{align}
  {\cal X}'_3 =&\, \frac{1}{3} {\cal X}_3^2 - {\cal I}_3 \nonumber \\ =&\,
  \sum_{\pm} \prod_{m=1}^{k} \delta_{i_m,i_{m\pm 1}'}, \nonumber \\
  {\cal S}'_3 =&\, \pm \left( {\cal S}_3^2 - q {\cal S}_3 \right) \nonumber \\ =&\,
  \sum_{\substack{m, m' = 1 \\ m \neq m'}}^{k} Z_{i_{m},i_{m+1}} Z^*_{i'_{m'},i'_{m'+1}} \delta_{i_{m+2},i'_{m'+2}},  
\end{align}
where the indices containing $m$ and $m'$ should be taken $\mod 3$. All five tensors ${\cal I}_3$, ${\cal S}_3$, ${\cal X}_3$, ${\cal S}_3'$, and ${\cal X}_3'$ are shown diagrammatically in Fig.\ \ref{fig:combined_diagrams}. The nontrivial multiplication rules are summarized in Table \ref{tab:mult3}.

Analogous to Eq.\ (\ref{eq:M2def}), for $k=3$ we introduce the $5 \times 5$ matrix $M_3$ such that
\begin{equation}
  \label{eq:M3def}
  {\cal L}_3 \begin{pmatrix} {\cal I}_3 \\ {\cal S}_3 \\ {\cal X}_3 \\ {\cal S}'_3 \\ {\cal X}'_3 \end{pmatrix}
  = M_3 \begin{pmatrix} {\cal I}_3 \\ {\cal S}_3 \\ {\cal X}_3 \\ {\cal S}'_3 \\{\cal X}'_3 \end{pmatrix}.
\end{equation}
{}From the multiplication rules of Tab.\ \ref{tab:mult3} we then find
\begin{align}
  M_3 = \begin{pmatrix}
  -\frac{3(q\mp 1)}{2} \!\! & 1 & -1 & 0 & 0 \\
  0 & -\frac{(q\mp 1)}{2}\!\! & 0 & 0 & 0 \\
  - 3 & \pm1 & -\frac{3(q\mp 1)}{2} \!\! & \pm1 & -3 \\
  0 & 0 & 0 & -\frac{(q\mp 1)}{2}\!\! & 0 \\
  0 & 0 & -2 & 1 & -\frac{3(q\mp 1)}{2}
\end{pmatrix}.
  \label{eq:M3}
\end{align}
Calculating ${\cal U}_3(t) = e^{{\cal L}_3 t}$ with the help of Eqs.\ (\ref{eq:M3def}) and (\ref{eq:M3}), we then find
\begin{align}
  \label{eq:U3}
  {\cal U}_3(t) =&\,
  \frac{1}{6}
  e^{-\frac{3}{2}(q-1)t} 
  \left[ \vphantom{\frac{M}{M}}
  (e^{-3 t}  + 4 + e^{3 t}) {\cal I}_3
  \right. \nonumber \\ &\, \left. \mbox{}
  - 2 \left( \frac{e^{-3 t}}{q+2} + \frac{2}{q-1} 
  - \frac{3 (q+1) e^{(q-1)t}}{(q+2)(q-1)} \right) {\cal S}_3
  \right. \nonumber \\ &\, \left. \mbox{}
  + 2 \left( \frac{-e^{-3 t}}{q+2} + \frac{1}{q-1} 
  - \frac{3 e^{(q-1)t}}{(q+2)(q-1)} \right) {\cal S}'_3
  \right. \nonumber \\ &\, \left. \mbox{}
  + (e^{-3 t} - e^{3 t}) {\cal X}_3
  + (e^{-3 t} - 2 + e^{3 t}) {\cal X}'_3
  \vphantom{\frac{M}{M}}
  \right].
\end{align}
for the orthogonal case.
The moment operator ${\cal U}_3(t)$ for the symplectic cases is obtained related by the duality transformation $t \to -t$, $q \to -q$, ${\cal S}_3 \to -{\cal S}_3$, ${\cal X}_3 \to -{\cal X}_3$,  ${\cal S}'_3 \to -{\cal S}'_3$.


\subsection{The case $k=4$}

The calculation of ${\cal U}_4(t)$ proceeds in the same way as that of $k=3$. In this case, the relevant algebra of rank-$8$ tensors has $12$ basis elements. Diagrammatic expressions for these tensors and their
multiplication rules are given in App. \ref{app:1}, as well as an expression for the matrix representing the action of ${\cal L}_4$ on the algebra and the moment operator ${\cal U}_4(t) = e^{{\cal L}_4 t}$.

\subsection{Trace moments}

The information about the moments of the evolution operator $U(t)$ can also be encoded in the trace moments 
\begin{equation}
  R_{i_1,\ldots,i_m} = \left\langle \prod_{j=1}^{m} \left( \frac{1}{q} \mbox{tr}\, U(t)^{i_j} \right) \right\rangle.
\end{equation}
With the help of Eqs.\ (\ref{eq:U1}), (\ref{eq:U2}), and (\ref{eq:U3}), we can obtain explicit expressions for the trace moments for $k = i_1 + \ldots + i_m \le 3$,
\begin{widetext}
\begin{align}
  \label{eq:R}
  \mathcal{R}_{1} =&\, e^{-\frac{1}{2}(q \mp 1) t}, \nonumber \\
\mathcal{R}_{1,1} =&\, \frac{e^{-(q \mp 1)t}}{2} \left( \frac{2}{q^2} e^{(q\mp 1)t} + \frac{q \mp 1}{q} e^{\pm t} + \frac{(q \mp 1)(q \pm 2)}{q^2} e^{\mp t} \right), \nonumber \\
\mathcal{R}_{2} =&\, \frac{e^{-(q \mp 1)t}}{2} \left[ \frac{\pm 2}{q} e^{(q\mp 1)t} + \frac{(q \mp 1)(2 \pm q)}{q} e^{\mp t} \mp (q \mp 1) e^{\pm t} \right], \nonumber \\
\mathcal{R}_{1,1,1} =&\, \frac{e^{-\frac{3}{2} (q \mp 1) t}}{3} \left[ \frac{2(q^2 - 4)}{q^2} + \frac{9}{q^2} e^{(q \mp 1) t} + \frac{q^2 \pm 3q - 4}{2q^2} e^{\mp 3t} + \frac{q^2 \mp 3q + 2}{2q^2} e^{\pm 3t} \right], \nonumber \\
\mathcal{R}_{1,2} =&\, \frac{e^{-\frac{3}{2} (q \mp 1) t}}{3} \left[ \frac{\pm 3}{q} e^{(q \mp 1) t} \pm \frac{q^2 \pm 3q - 4}{2q} e^{\mp 3t} \mp \frac{q^2 \mp q + 2}{2q} e^{\pm 3t} \right], \nonumber \\
\mathcal{R}_{3} =&\, \frac{e^{-\frac{3}{2} (q \mp 1) t}}{3} \left[ 4 - q^2 + \frac{q^2 \pm 3q - 4}{2} e^{\mp 3t} + \frac{q^2 \mp 3q + 2}{2} e^{\pm 3t} \right].
\end{align}
\end{widetext}
Expressions for $k=4$ are given in the appendix. 

\section{Brownian ensembles over ${\rm SO}^{-}(q)$}
\label{sec:4}

Since the sign of the determinant $\det U$ of an orthogonal matrix $U$ cannot change continuously, the orthogonal group ${\rm O}(q)$ consists of two disconnected components. These are the special orthogonal group ${\rm SO}(q)$, which contains $q \times q$ matrices $U$ with $\det U = 1$, and the remainder ${\rm SO}^-(q)$, which contains matrices $U$ with $\det U = -1$.
The Brownian process of the previous section generates matrices along a continuous path from the identity. Hence, for the orthogonal case it explores only the identity component $\mathrm{SO}(q)$ of the orthogonal group, relaxing from $\openone$ to the Haar measure on $\mathrm{SO}(q)$. This process never reaches the second component $\mathrm{SO}^-(q)$.

We here show how the Brownian ensemble can nevertheless be used to build an interpolation ensemble with $\mathrm{SO}^-(q)$ that relaxes from a ``minimal'' ensemble defined on $\mathrm{SO}^-(q)$ at $t=0$ to the uniform Haar-distributed ensemble in the limit $t \to \infty$. Hereto, we consider the Brownian evolution with a nontrivial ensemble of evolution operators $V \equiv U(0)$ at time $t=0$, {\em i.e.}, 
\begin{equation}
  U(t) = V \tilde U(t),\ \
  \tilde U(t) = {\cal T}_t e^{-i \int_0^t dt' H(t')}.
\end{equation}
The  moment operator ${\cal U}_k(t)$ for this ensemble is
\begin{equation}
  {\cal U}_k(t) = {\cal V}_k \tilde {\cal U}_k(t),
\end{equation}
where ${\cal U}_k(t)$ is given by the expressions in Sec.\ \ref{sec:2} for $k \le 4$ and ${\cal V}$ depends on the choice of the initial condition. 

A simple choice for the initial condition $V$ is
\begin{equation}
  \label{eq:V}
  V = O\, \mbox{diag}(-1,1,\ldots,1) O^{\rm T},
\end{equation}
with $O$ a Haar-distributed special orthogonal matrix. (For simplicity, we here have taken the standard choice $Z = 1$ for the involution matrix.) 
This choice for $V$ may be considered ``minimal'' in the sense that $V$ constitutes the slowest-scrambling conjugation-invariant ensemble in the $\mathrm{SO}^-(q)$ sector in the context of operator spreading speed \cite{tan2026}. Further, in quantum information, $-V$ can be interpreted as the reflection-about-a-state primitive underlying Grover search and Szegedy quantum walks \cite{Grover1996,Szegedy2004}, evaluated with a Haar-random axis. In condensed-matter settings, $V$ can be interpreted as a random single impurity or blocked channel among $q$ otherwise transparent ones; for $q=3$ it realizes a $\pi$-pulse about a random axis, and for $q=2$ a random mirror.

The  moment operators ${\cal V}_k$ can be easily calculated from the trace moments,
\begin{align}
  \label{eq:V1}
  {\cal V}_1 =&\, \frac{q-2}{q} {\cal I}_1, \nonumber \\
  {\cal V}_2 =&\, \frac{1}{q(q+2)} \left[
  (q^2-2q-4) {\cal I}_2 + 4 {\cal X}_2 + 4 {\cal S}_2 \right], \nonumber \\
  {\cal V}_3 =&\ \frac{1}{q(q+2)(q+4)}[(q^3-16q-8)\mathcal{I}_3-8(\mathcal{S}'_3+\mathcal{X}'_3)\nonumber\\
  &\, +4(q+2)(\mathcal{X}_3+\mathcal{S}_3)].
\end{align}
The expression for ${\cal V}_k$ for $k=4$ and a framework to calculate ${\cal V}_k$ for general $k$ can be found in App. \ref{app:2}.
From here, the full moment operators ${\cal U}_k(t)$ can be easily calculated using the multiplication rules for the rank-$2 k$ tensors appearing in these expressions, see Sec.\ \ref{sec:2}.

The Brownian ensemble constructed this way is supported entirely on $\mathrm{SO}^-(q)$, has an orthogonal-invariant probability distribution for all times $t$, and relaxes to the Haar measure on $\mathrm{SO}^-(q)$ in the long-time limit. The condition that the interpolating ensemble be orthogonal invariant rules out seeding the diffusion from a fixed matrix, in contrast to the case of $\mathrm{SO}(q)$, where the unit matrix is an orthogonal-invariant starting point of the Brownian ensemble. Other initial conditions, such as those starting from a matrix $V = U(0)$ with a larger (but still odd) number of negative eigenvalues can be constructed in the same manner.

\section{Applications}
\label{sec:3}

The moment operators and the trace moments can be used for applications in the fields of quantum chaos, quantum information, and quantum transport. 

\subsection{Frame potentials}

One example in the field of quantum information is the frame potential $F^{(k)}(t)$, which measures how close the Brownian ensemble is to forming an orthogonal/symplectic $k$-design \cite{roberts_chaos_2017,shaya2026}. The frame potential $F^{(k)}(t)$ is defined as the ensemble average
\begin{equation}
  F^{(k)}(t)= \langle \left( \mbox{tr}\, U(t) V(t)^{\dagger} \right)^{2k} \rangle,
\end{equation}
where the average is taken over independently distributed matrices $U(t)$ and $V(t)$ from the Brownian ensemble. This average can be expressed in terms of the moment operator as
\begin{align}
  F^{(k)}(t) =&\, \tr \left( \mathcal{U}_{k}(t) \mathcal{U}_{k}(t) \right) 
  \nonumber \\ =&\, \tr e^{2\mathcal{L}_{k} t}.
\end{align}
Consequently, the first and second frame potentials are directly determined by the trace moments as:
\begin{align}
  F^{(1)}(t) &= q^2\mathcal{R}_{1,1}(2t), \\
  F^{(2)}(t) &= q^4 \mathcal{R}_{1,1;1,1}(2t).
\end{align}
By extracting the slowest decaying transient terms from $F^{(1)}(t)$ and $F^{(2)}(t)$, we find that, after rescaling to the physical time ($t \to t' = qt$), the characteristic scrambling time $t'_{\rm scr}$ for convergence to the Haar measure scales as ($\sim \log q$), consistent with the results in the existing literature for the unitary case \cite{jian2023linear,tang2024brownian}. 

Other applications of the trace moments in the field of quantum information are the out-of-time-order correlators~\cite{hosur2016chaos,roberts_chaos_2017} and the second R\'enyi entropy~\cite{hosur2016chaos,nahum2017growth}. At the level of the two-replica diagrammatic expansion, the same machinery yields the shadow norm that controls the sample complexity of classical shadow tomography~\cite{huang2020shadows}. In Refs.\ \cite{tan2025operatorspreading,tan2026}, we use the trace moments of the two-qudit gate operator to calculate the butterfly velocity and diffusion constant describing operator spreading in a random unitary circuit with brickwork architecture \cite{Fisher2023random}. 

\subsection{Spectral form factor}

The trace moments also allow us to directly evaluate the spectral form factor. The spectral form factor $K(t)$ characterizes the correlations between pairs of energy eigenvalues in the spectrum, diagnosing the presence of level repulsion \cite{mehta1997random,stoeckmann1999,guhr1998,cotler2017chaos,cotler2017black}. It is given directly by the first trace moment,
\begin{equation}
    K(t) = q^2\mathcal{R}_{1,1}(t),
\end{equation}
which coincides with the first frame potential up to a factor of two in the time argument, such that $F^{(1)}(t) = K(2t)$. The variance of the spectral form factor can also be expressed in terms of the trace moments \cite{prange1997sff},
\begin{equation}
  \mbox{var}\, K(t) = q^4\mathcal{R}_{1,1;1,1}(t)-q^4\mathcal{R}^2_{1,1}(t) .
\end{equation}

\subsection{Heat conductance}

\begin{figure}
\includegraphics[scale=0.65,trim={ 4.5cm 20cm 4.5cm 2.5cm},clip ]{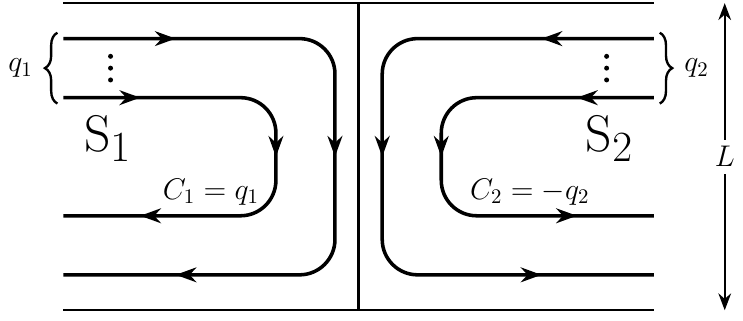}
\caption{\justifying \small \label{fig:junction}
Junction between topological superconductors ${\rm S}_1$ and ${\rm S}_2$ with Chern numbers $C_1 > 0$ and $C_2 < 0$. The number of chiral Majorana edge modes at the superconductor-vacuum boundary is $q_j = |C_j|$, $j=1,2$. At the interface between the two superconductors, there are $q = q_1 + q_2$ co-propagating chiral Majorana modes. }
\end{figure}

As an application in the field of quantum transport, we consider scattering between co-propagating edge modes of a topological superconductor in a junction connecting topological phases with Chern numbers of opposite sign \cite{serban2010,dahlhaus2010}, see Fig.\ \ref{fig:junction}. (The Chern numbers are defined with respect to the Bogoliubov-de Gennes Hamiltonian of the superconductors.) If spin-rotation symmetry and time-reversal symmetry are both broken, which corresponds to symmetry class D from the Tenfold-way classification \cite{altland1997}, such a junction has a heat conductance that depends on the degree of scattering between the co-propagating modes at the junction. 

We consider superconductors ${\rm S}_1$ and ${\rm S}_2$ with Chern numbers $C_1 = q_1$ and $C_2 = - q_2$, where $q_1$ and $q_2$ are both positive. These have $q_1$ and $q_2$ chiral Majorana modes at the boundary with the vacuum and $q = q_1 + q_2$ co-propagating Majorana modes along the interface between the two superconductors, see Fig.\ \ref{fig:junction}. Scattering between the edge modes is described by the $q \times q$ scattering matrix $U_{ij}$, where $i,j=1,\ldots,q_1$ is associated with ${\rm S}_1$ and $i,j=q_1+1,\ldots,q$ with ${\rm S}_2$. A temperature difference between ${\rm S}_1$ and ${\rm S}_2$ results in a heat flow, whereby the edge-state contribution to the heat conductance $G_{Q}$ is given by \cite{dahlhaus2010}
\begin{equation}
  G_{Q} = G_{Q,0} 
  \sum_{i=1}^{q_1} \sum_{j=q_1+1}^{q} |U_{ij}|^2,
  \label{eq:GQ}
\end{equation}
with $G_{Q,0} = {\pi^2 k_{\rm B}^2 T}/{6 h}$.

In the special case that the co-propagating Majorana modes at the interface do not mix in the absence of disorder, which is the case, {\em e.g.}, if the transition between the two superconducting phases is smooth, the scattering matrix ensemble is given by the Brownian motion ensemble (\ref{eq:brownian0}) for orthogonal matrices, whereby the Brownian time $t$ is proportional to $L/l$, $L$ being the junction length and $l$ the mean free path. Hence, for the average and variance of the heat conductance we find
\begin{align}
  \langle G_{Q}(t) \rangle =&\,\frac{q_1 q_2}{q} G_{Q,0} (1-e^{-qt}) , \\
  \mbox{var}\, G_{Q}(t) =&\, 
  q_1 q_2 G_{Q,0}^2 \left\{ 
  \vphantom{\frac{M}{M}}
  \right. \left. 
  \left[
  (C_{\cal \tilde S}(t)-C_{\cal \tilde S}(\infty))(q_1 q_2+2)
  \right. \right. \nonumber \\ &\, \ \ \ \ \ \ \ \ 
  \left. \left. \mbox{}
  + 2 (C_{\cal \tilde S'}(t)-C_{\cal \tilde S'}(\infty)) (q+1) \right] \right. \nonumber\\ 
  &\, \left. \mbox{} + \frac{q_1q_2}{q^2} 
  \left[ \frac{2}{(q-1)(q+2)}   
+ 2 e^{-q t} - e^{-2 q t}
  \right] \right\} , \nonumber
\end{align} 
where $C_{\cal \tilde S}(t)$ and $C_{\cal \tilde S'}(t)$ are defined in Appendix \ref{app:1}, with $C_{\cal \tilde S}(\infty)$ and $C_{\cal \tilde S'}(\infty)$ representing their corresponding infinite-time limits. 
In the limit $t \to \infty$ one recovers results from the circular real ensemble \cite{dahlhaus2010,beenakker2011}. If there is mixing between the co-propagating Majorana modes already in the absence of disorder, one can still use a description in terms of a Brownian motion ensemble, but with a nontrivial starting point $U(0)$ at $t=0$. The disorder average is then calculated using the method of Sec.\ \ref{sec:4}.

Analogously, the Brownian motion ensemble in the symplectic group can be used to describe disorder scattering between co-propagating chiral modes at the interface between two topological superconductors with spin-rotation symmetry. These are in class C of the Tenfold-way classification. In this case, the heat conductivity is still given by Eq.\ (\ref{eq:GQ}), but with $G_{Q,0} = {\pi^2 k_{\rm B}^2 T}/{3 h}$ \cite{sivan1986,butcher1990}.

\section{Conclusion and Outlook}
\label{sec:5}

In this work, we developed a diagrammatic framework to calculate polynomial averages for Brownian ensembles of orthogonal and symplectic matrices. The combinatorial backbone of this framework is the Brauer algebra~\cite{braueralgebra1937,wenzl1988,collins2006}. We enumerated all of its diagrams up to $k=4$, organized them into equivalence classes, and constructed the corresponding multiplication tables, a procedure that can be extended systematically to higher $k$. Building on this, we computed the moment operators and trace moments up to fourth order, and related them to two physical diagnostics, the frame potential and the spectral form factor. These diagnostics reveal that, after rescaling to the physical time ($t \to t' = qt$), the convergence time to reach an orthogonal or symplectic 2-design scales as $\sim \log q$. We also considered an application to disorder-induced scattering between co-propagating chiral Majorana modes, which appear, {\em e.g.}, at a junction between topological superconductors \cite{serban2010,dahlhaus2010}.

Recognizing that the continuous Brownian process is topologically confined to the special orthogonal group $\mbox{SO}(q)$ if the starting point is the $q \times q$ identity matrix, we used a nontrivial initial condition $U(0)$ to construct an interpolation ensemble that accesses the disconnected sector $\mathrm{SO}^-(q)$ and interpolates between a ``minimal'' distribution and the fully random Haar distribution. Crucially, this interpolating ensemble remains strictly orthogonally invariant at every time $t$, rather than merely in the asymptotic limit. 
Since the moment operators are multiplicative, our results can easily be generalized to other Brownian ensembles with a nontrivial initial condition.


The ensembles constructed here provide a controlled setting for dynamical phase transitions. Interspersing the Brownian evolution with projective measurements would enable an analytic study of measurement-induced entanglement transitions~\cite{skinner2019mipt,li2018zeno,li2019measurement}, a direction in which closely related solvable Brownian models have proven especially tractable~~\cite{jian2021syk,bentsen2021purification,sahu2022longrange,gerbino2024dyson}. Incorporating decoherence channels opens a parallel route to noise-induced phase transitions~\cite{liu2024noise}. Together, these would extend the framework from static spectral diagnostics to the dynamics of monitored and open quantum systems.

\begin{acknowledgements}
Special thanks go to Yiping Deng for suggestions on figure editing. Financial support was provided by the Einstein Stiftung Berlin (Einstein Research Unit on Quantum Devices) and by the Deutsche Forschungsgemeinschaft (DFG, German Research Foundation) - Project Number 277101999 - CRC TR 183 (project A03).

\end{acknowledgements}

\appendix

\section{Diagrams and multiplication rule for $k=4$}
\label{app:1}

\begin{figure*}[t!]
    \centering
 \includegraphics[trim=2cm 1.7cm 0.5cm 1cm, clip, scale=0.8]{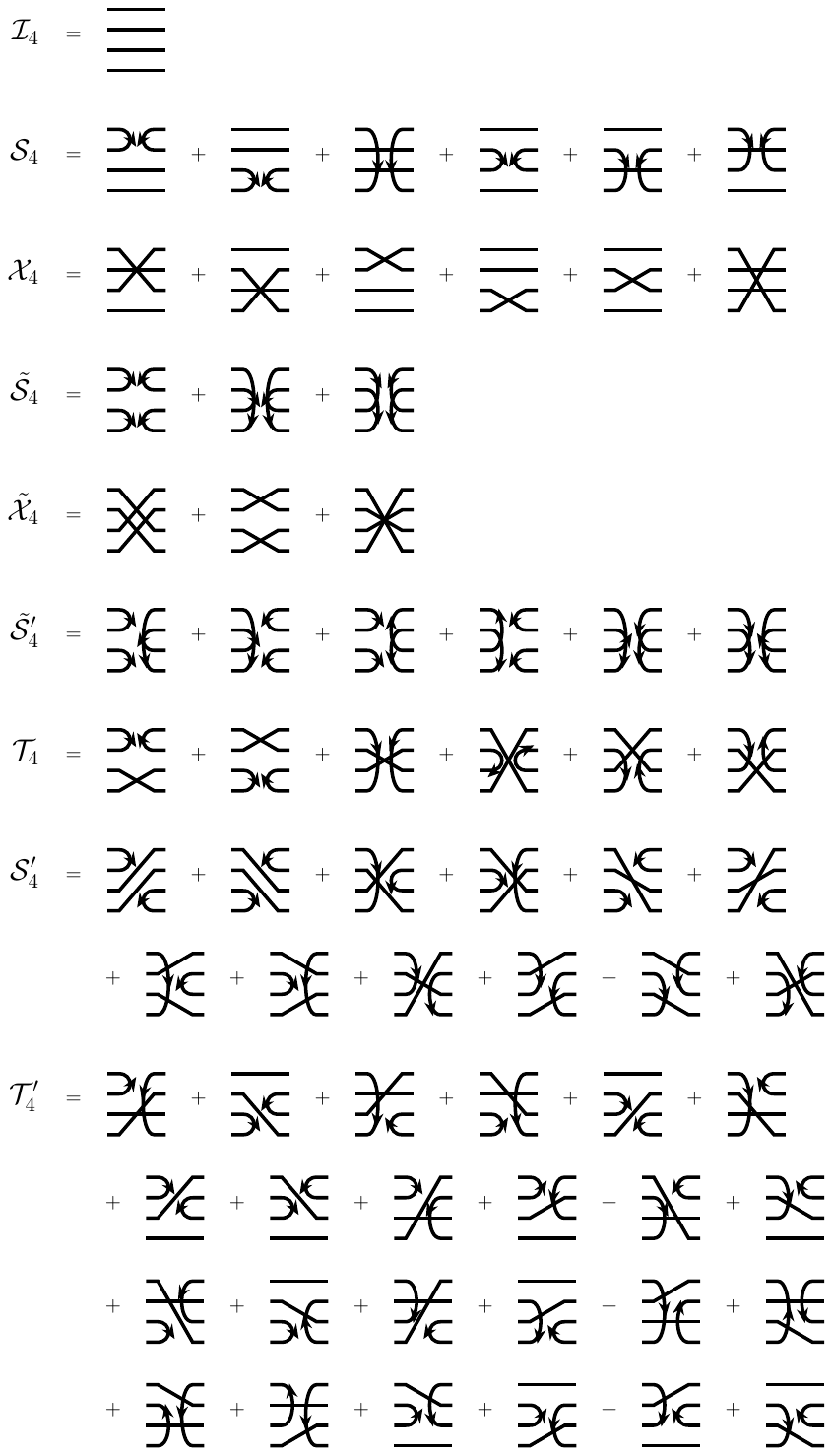}
\vspace{0cm}
\end{figure*}
\begin{figure*}[t!]
    \centering
\includegraphics[trim=2cm 15cm 0.5cm .5cm, clip, scale=0.8]{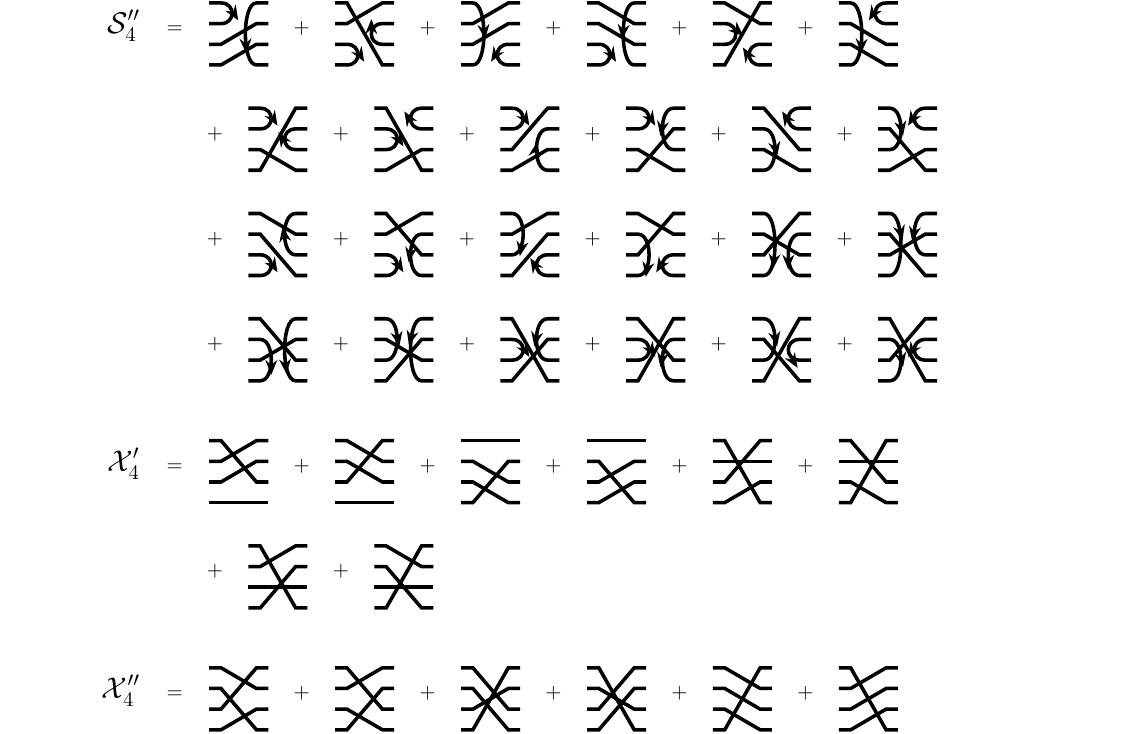}
\caption{\justifying \small  Diagrammatic representations of ${\cal I}_4$, ${\cal S}_4$, ${\cal X}_4$, $\tilde {\cal S}_4$, $\tilde {\cal X}_4$, $\tilde {\cal S}'_4$, ${\cal T}_4$, ${\cal S}'_4$, ${\cal T}'_4$, ${\cal S}''_4$, ${\cal X}'_4$, and ${\cal X}''_4$ for the symplectic case. Diagrammatic representations for the orthogonal case are obtained by omitting the orientations on the arcs connecting indices on the same side of each diagram. }
    \label{fig:symplectic}
\end{figure*}

In this appendix, we present the results for the calculation of the moment operator ${\cal U}_4(t)$ for $k=4$. In this case, the relevant Brauer algebra is spanned by 12 linearly independent tensors. In addition to the three tensors $\mathcal{I}_4$, $\mathcal{S}_4$, and $\mathcal{X}_4$ introduced in Sec.\ \ref{sec:2}, we denote these basis tensors as $ \tilde{\mathcal{S}}_4$, $ \tilde{\mathcal{X}}_4$, $ \tilde{\mathcal{S}}'_4$, $ \mathcal{T}_4$, $ \mathcal{S}'_4$, $ \mathcal{T}'_4$, $ \mathcal{S}''_4$, $ \mathcal{X}'_4$, and $ \mathcal{X}''_4$ . Their diagrammatic representation for the symplectic case is shown in Fig.\ \ref{fig:symplectic}. The diagrammatic representation for the orthogonal case is obtained by omitting the orientation of solid lines connecting indices on the same side of the diagrams. 

To find the multiplication rules, we note that for the symplectic diagrams, arc orientations are governed strictly by topological crossing parity. Assuming a default downward flow, each line intersection contributes a factor $-1$. Consequently, diagrams with an even number of crossings retain a purely downward orientation, whereas those with an odd number require exactly one arc to be inverted upward to absorb the net negative sign and ensure algebraic consistency. 

The algebra spanned by $\mathcal{I}_4$, $\mathcal{S}_4$, $ \mathcal{X}_4$, $ \tilde{\mathcal{S}}_4$, $ \tilde{\mathcal{X}}_4$, $ \tilde{\mathcal{S}}'_4$, $ \mathcal{T}_4$, $ \mathcal{S}'_4$, $ \mathcal{T}'_4$, $ \mathcal{S}''_4$, $ \mathcal{X}'_4$, and $ \mathcal{X}''_4$ is commutative. Multiplication by $\mathcal{I}_4$ is trivial. The nontrivial multiplication rules are (omitting the subscript $k$ with $k=4$ for symplicity of notation)
\begin{align*}
\mathcal{S} \ttimes \mathcal{S} &= \NN\mathcal{S} + 2\tilde{\mathcal{S}} \pm \mathcal{T}' \\
\mathcal{S} \ttimes \mathcal{X} &= \pm (\mathcal{S} + \mathcal{T}' + \mathcal{T}) \\
\mathcal{S} \ttimes \tilde{\mathcal{S}} &= 2\NN\tilde{\mathcal{S}} \pm 2\tilde{\mathcal{S}}' \\
\mathcal{S} \ttimes \tilde{\mathcal{X}} &= \mathcal{T} + \mathcal{S}' \\
\mathcal{S} \ttimes \tilde{\mathcal{S}}' &= 2\NN\tilde{\mathcal{S}}' \pm (2\tilde{\mathcal{S}}' + 4\tilde{\mathcal{S}}) \\
\mathcal{S} \ttimes \mathcal{T} &= \NN\mathcal{T} + 2\tilde{\mathcal{S}} \pm \mathcal{S}'' \\
\mathcal{S} \ttimes \mathcal{S}' &= \NN\mathcal{S}' + 4\tilde{\mathcal{S}} \pm (\mathcal{T}' + \mathcal{S}'') \\
\mathcal{S} \ttimes \mathcal{T}' &= \NN\mathcal{T}' + 4\tilde{\mathcal{S}}' \pm (4\mathcal{S} + 2\mathcal{S}' + \mathcal{T}' + \mathcal{S}'') \\
\mathcal{S} \ttimes \mathcal{S}'' &= \NN\mathcal{S}'' + 4\tilde{\mathcal{S}}' \pm (4\mathcal{T} + 2\mathcal{S}' + \mathcal{T}' + \mathcal{S}'') \\
\mathcal{S} \ttimes \mathcal{X}' &= \mathcal{T}' + \mathcal{S}'' \\
\mathcal{S} \ttimes \mathcal{X}'' &= \pm (\mathcal{S}' + \mathcal{S}'')
  , \\
%
\mathcal{X} \ttimes \mathcal{X} &= 6\mathcal{I} + 2\tilde{\mathcal{X}} + 3\mathcal{X}' \\
\mathcal{X} \ttimes \tilde{\mathcal{S}} &= \pm (2\tilde{\mathcal{S}} + 2\tilde{\mathcal{S}}') \\
\mathcal{X} \ttimes \tilde{\mathcal{X}} &= \mathcal{X} + 2\mathcal{X}'' \\
\mathcal{X} \ttimes \tilde{\mathcal{S}}' &= \pm (4\tilde{\mathcal{S}}' + 4\tilde{\mathcal{S}}) \\
\mathcal{X} \ttimes \mathcal{T} &= \pm (\mathcal{T} + \mathcal{S} + \mathcal{S}'') \\
\mathcal{X} \ttimes \mathcal{S}' &= \pm (2\mathcal{S}' + \mathcal{T}' + \mathcal{S}'') \\
\mathcal{X} \ttimes \mathcal{T}' &= \pm (2\mathcal{S}'' + 4\mathcal{S} + 2\mathcal{S}' + 2\mathcal{T}') \\
\mathcal{X} \ttimes \mathcal{S}'' &= \pm (2\mathcal{S}'' + 4\mathcal{T} + 2\mathcal{S}' + 2\mathcal{T}') \\
\mathcal{X} \ttimes \mathcal{X}' &= 4\mathcal{X} + 4\mathcal{X}'' \\
\mathcal{X} \ttimes \mathcal{X}'' &= 4\tilde{\mathcal{X}} + 3\mathcal{X}'
, \\
%
\tilde{\mathcal{S}} \ttimes \tilde{\mathcal{S}} &= \NN^2\tilde{\mathcal{S}} \pm \NN\tilde{\mathcal{S}}' \\
\tilde{\mathcal{S}} \ttimes \tilde{\mathcal{X}} &= 3\tilde{\mathcal{S}} \\
\tilde{\mathcal{S}} \ttimes \tilde{\mathcal{S}}' &= \NN^2\tilde{\mathcal{S}}' \pm (2\NN\tilde{\mathcal{S}} + \NN\tilde{\mathcal{S}}') \\
\tilde{\mathcal{S}} \ttimes \mathcal{T} &= 2\NN\tilde{\mathcal{S}} \pm 2\tilde{\mathcal{S}}' \\
\tilde{\mathcal{S}} \ttimes \mathcal{S}' &= 4\NN\tilde{\mathcal{S}} \pm 4\tilde{\mathcal{S}}' \\
\tilde{\mathcal{S}} \ttimes \mathcal{T}' &= 4\NN\tilde{\mathcal{S}}' \pm (4\tilde{\mathcal{S}}' + 8\tilde{\mathcal{S}}) \\
\tilde{\mathcal{S}} \ttimes \mathcal{S}'' &= 4\NN\tilde{\mathcal{S}}' \pm (4\tilde{\mathcal{S}}' + 8\tilde{\mathcal{S}}) \\
\tilde{\mathcal{S}} \ttimes \mathcal{X}' &= 4\tilde{\mathcal{S}}' \\
\tilde{\mathcal{S}} \ttimes \mathcal{X}'' &= \pm (2\tilde{\mathcal{S}} + 2\tilde{\mathcal{S}}')
, \\
%
\tilde{\mathcal{X}} \ttimes \tilde{\mathcal{X}} &= 3\mathcal{I} + 2\tilde{\mathcal{X}} \\
\tilde{\mathcal{X}} \ttimes \tilde{\mathcal{S}}' &= 3\tilde{\mathcal{S}}' \\
\tilde{\mathcal{X}} \ttimes \mathcal{T} &= \mathcal{S}' + \mathcal{S} \\
\tilde{\mathcal{X}} \ttimes \mathcal{S}' &= 2\mathcal{S} + 2\mathcal{T} + \mathcal{S}' \\
\tilde{\mathcal{X}} \ttimes \mathcal{T}' &= \mathcal{T}' + 2\mathcal{S}'' \\
\tilde{\mathcal{X}} \ttimes \mathcal{S}'' &= 2\mathcal{T}' + \mathcal{S}'' \\
\tilde{\mathcal{X}} \ttimes \mathcal{X}' &= 3\mathcal{X}' \\
\tilde{\mathcal{X}} \ttimes \mathcal{X}'' &= 2\mathcal{X} + \mathcal{X}''
, \\
%
\tilde{\mathcal{S}}' \ttimes \tilde{\mathcal{S}}' &= \NN^2\tilde{\mathcal{S}}' + 2\NN^2\tilde{\mathcal{S}} \pm (2\tilde{\mathcal{S}} + 3\NN\tilde{\mathcal{S}}') \\
\tilde{\mathcal{S}}' \ttimes \mathcal{T} &= 2\NN\tilde{\mathcal{S}}' \pm (2\tilde{\mathcal{S}}' + 4\tilde{\mathcal{S}}) \\
\tilde{\mathcal{S}}' \ttimes \mathcal{S}' &= 4\NN\tilde{\mathcal{S}}' \pm (8\tilde{\mathcal{S}} + 4\tilde{\mathcal{S}}') \\
\tilde{\mathcal{S}}' \ttimes \mathcal{T}' &= 8\NN\tilde{\mathcal{S}} + 4\NN\tilde{\mathcal{S}}' \pm (8\tilde{\mathcal{S}} + 12\tilde{\mathcal{S}}') \\
\tilde{\mathcal{S}}' \ttimes \mathcal{S}'' &= 8\NN\tilde{\mathcal{S}} + 4\NN\tilde{\mathcal{S}}' \pm (8\tilde{\mathcal{S}} + 12\tilde{\mathcal{S}}') \\
\tilde{\mathcal{S}}' \ttimes \mathcal{X}' &= 8\tilde{\mathcal{S}} + 4\tilde{\mathcal{S}}' \\
\tilde{\mathcal{S}}' \ttimes \mathcal{X}'' &= \pm (4\tilde{\mathcal{S}} + 4\tilde{\mathcal{S}}')
, \\
%
\mathcal{T} \ttimes \mathcal{T} &= \NN\mathcal{S} + 2\tilde{\mathcal{S}} \pm \mathcal{T}' \\
\mathcal{T} \ttimes \mathcal{S}' &= \NN\mathcal{S}' + 4\tilde{\mathcal{S}} \pm (\mathcal{T}' + \mathcal{S}'') \\
\mathcal{T} \ttimes \mathcal{T}' &= \NN\mathcal{S}'' + 4\tilde{\mathcal{S}}' \pm (4\mathcal{T} + 2\mathcal{S}' + \mathcal{S}'' + \mathcal{T}') \\
\mathcal{T} \ttimes \mathcal{S}'' &= \NN\mathcal{T}' + 4\tilde{\mathcal{S}}' \pm (4\mathcal{S} + 2\mathcal{S}' + \mathcal{S}'' + \mathcal{T}') \\
\mathcal{T} \ttimes \mathcal{X}' &= \mathcal{T}' + \mathcal{S}'' \\
\mathcal{T} \ttimes \mathcal{X}'' &= \pm (\mathcal{S}' + \mathcal{T}')
, \\
%
\mathcal{S}' \ttimes \mathcal{S}' &= 2\NN\mathcal{S} + 8\tilde{\mathcal{S}} + 2\NN\mathcal{T} \pm (2\mathcal{T}' + 2\mathcal{S}'') \\
\mathcal{S}' \ttimes \mathcal{T}' &= \NN\mathcal{T}' + \NN\mathcal{S}'' + 8\tilde{\mathcal{S}}' \\ 
&\quad \pm (4\mathcal{S} + 4\mathcal{T} + 4\mathcal{S}' + 2\mathcal{T}' + 2\mathcal{S}'') \\
\mathcal{S}' \ttimes \mathcal{S}'' &= \NN\mathcal{T}' + \NN\mathcal{S}'' + 8\tilde{\mathcal{S}}' \\ 
&\quad \pm (4\mathcal{S} + 4\mathcal{T} + 4\mathcal{S}' + 2\mathcal{T}' + 2\mathcal{S}'') \\
\mathcal{S}' \ttimes \mathcal{X}' &= 2\mathcal{T}' + 2\mathcal{S}'' \\
\mathcal{S}' \ttimes \mathcal{X}'' &= \pm (2\mathcal{S} + 2\mathcal{T} + \mathcal{T}' + \mathcal{S}'')
, \\
%
\mathcal{T}' \ttimes \mathcal{T}' &= 4\NN\mathcal{S} + 16\tilde{\mathcal{S}} + 2\NN\mathcal{S}' + \NN\mathcal{T}' + \NN\mathcal{S}'' \\
&\quad \pm (4\mathcal{S} + 8\tilde{\mathcal{S}}' + 4\mathcal{T} + 4\mathcal{S}' + 8\mathcal{T}' + 4\mathcal{S}'') \\
\mathcal{T}' \ttimes \mathcal{S}'' &= 4\NN\mathcal{T} + 16\tilde{\mathcal{S}} + 2\NN\mathcal{S}' + \NN\mathcal{T}' + \NN\mathcal{S}'' \\
&\quad \pm (4\mathcal{S} + 8\tilde{\mathcal{S}}' + 4\mathcal{T} + 4\mathcal{S}' + 4\mathcal{T}' + 8\mathcal{S}'') \\
\mathcal{T}' \ttimes \mathcal{X}' &= 4\mathcal{S} + 4\mathcal{T} + 4\mathcal{S}' + 2\mathcal{T}' + 2\mathcal{S}'' \\
\mathcal{T}' \ttimes \mathcal{X}'' &= \pm (4\mathcal{T} + 2\mathcal{S}' + 2\mathcal{T}' + 2\mathcal{S}'')
, \\
%
\mathcal{S}'' \ttimes \mathcal{S}'' &= 4\NN\mathcal{S} + 16\tilde{\mathcal{S}} + 2\NN\mathcal{S}' + \NN\mathcal{T}' + \NN\mathcal{S}'' \\
&\quad \pm (4\mathcal{S} + 8\tilde{\mathcal{S}}' + 4\mathcal{T} + 4\mathcal{S}' + 8\mathcal{T}' + 4\mathcal{S}'') \\
\mathcal{S}'' \ttimes \mathcal{X}' &= 4\mathcal{S} + 4\mathcal{T} + 4\mathcal{S}' + 2\mathcal{T}' + 2\mathcal{S}'' \\
\mathcal{S}'' \ttimes \mathcal{X}'' &= \pm (4\mathcal{S} + 2\mathcal{S}' + 2\mathcal{T}' + 2\mathcal{S}'') \\
\mathcal{X}' \ttimes \mathcal{X}' &= 8\mathcal{I} + 8\tilde{\mathcal{X}} + 4\mathcal{X}' \\
\mathcal{X}' \ttimes \mathcal{X}'' &= 4\mathcal{X} + 4\mathcal{X}'' \\
\mathcal{X}'' \ttimes \mathcal{X}'' &= 6\mathcal{I} + 2\tilde{\mathcal{X}} + 3\mathcal{X}'
\end{align*}
The matrix $M_4$ representing the action of $\mathcal{L}_4$ on the algebra is 
\begin{widetext}

\setcounter{MaxMatrixCols}{12}
\begin{equation}
M_4 =\begin{pmatrix}
\pm 2 - 2q & 1 & -1 & 0 & 0 & 0 & 0 & 0 & 0 & 0 & 0 & 0 \\
0 & \pm 1 - q & 0 & 2 & 0 & 0 & \mp 1 & 0 & 0 & 0 & 0 & 0 \\
-6 & \pm 1 & \pm 2 - 2q & 0 & -2 & 0 & \pm 1 & 0 & \pm 1 & 0 & -3 & 0 \\
0 & 0 & 0 & 0 & 0 & 0 & 0 & 0 & 0 & 0 & 0 & 0 \\
0 & 0 & -1 & 0 & \pm 2 - 2q & 0 & 1 & 1 & 0 & 0 & 0 & -2 \\
0 & 0 & 0 & 0 & 0 & 0 & 0 & 0 & 0 & 0 & 0 & 0 \\
0 & \mp 1 & 0 & 2 & 0 & 0 & \pm 1 - q & 0 & 0 & 0 & 0 & 0 \\
0 & 0 & 0 & 4 & 0 & 0 & 0 & -q & 0 & 0 & 0 & 0 \\
0 & 0 & 0 & 0 & 0 & 4 & 0 & 0 & \pm 1 - q & \mp 1 & 0 & 0 \\
0 & 0 & 0 & 0 & 0 & 4 & 0 & 0 & \mp 1 & \pm 1 - q & 0 & 0 \\
0 & 0 & -4 & 0 & 0 & 0 & 0 & 0 & 1 & 1 & \pm 2 - 2q & -4 \\
0 & 0 & 0 & 0 & -4 & 0 & 0 & \pm 1 & 0 & \pm 1 & -3 & \pm 2 - 2q
\end{pmatrix},
\end{equation}
\end{widetext}
where the upper signs correspond to the orthogonal case and the lower signs correspond to the symplectic case. Finally, the resulting expression for the moment operator ${\cal U}_4(t) = e^{{\cal L}_4 t}$ is 
\begin{widetext}
\begin{align}
\mathcal{U}_4(t) =  &C_{\mathcal{I}}(t)\mathcal{I}_4 + C_{\mathcal{S}}(t)\mathcal{S}_4 + C_{\mathcal{X}}(t)\mathcal{X}_4 + C_{\tilde{\mathcal{S}}}(t)\tilde{\mathcal{S}}_4 + C_{\tilde{\mathcal{X}}}(t)\tilde{\mathcal{X}}_4 + C_{\tilde{\mathcal{S}}'}(t)\tilde{\mathcal{S}}'_4 \nonumber\\
&+ C_{\mathcal{T}}(t)\mathcal{T}_4 + C_{\mathcal{S}'}(t)\mathcal{S}'_4 + C_{\mathcal{T}'}(t)\mathcal{T}'_4 + C_{\mathcal{S}''}(t)\mathcal{S}''_4 + C_{\mathcal{X}'}(t)\mathcal{X}'_4 + C_{\mathcal{X}''}(t)\mathcal{X}''_4 ,
\end{align}
where the coefficients are defined as
\begin{align}
C_{\mathcal{I}}(t) &= \frac{e^{-2(2+q)t}}{24} \left( 1 + 9e^{4t} + 4e^{6t} + 9e^{8t} + e^{12t} \right), \\
C_{\mathcal{S}}(t) &= \frac{e^{-2(2+q)t}}{12(-2+q)q(2+q)(4+q)} \Big[ 6e^{6t+qt}q^2(4+q) - q(-4+q^2)  - 2e^{6t}q(8+6q+q^2) - 3e^{8t}q(8+6q+q^2) \nonumber \\
&\quad - 6e^{4t}(-8-6q+3q^2+q^3) + 6e^{4t+qt}(-8+4q^2+q^3) \Big], \\
C_{\mathcal{X}}(t) &= -\frac{e^{-2(2+q)t}}{24} \left( -1 - 3e^{4t} + 3e^{8t} + e^{12t} \right), \\
C_{\tilde{\mathcal{S}}}(t) &= \frac{e^{-2(2+q)t}}{3(-2+q)(-1+q)q(2+q)(4+q)} \Big[ -6e^{4t+qt}(-1+q)(2+q)^2  + q(2-3q+q^2) + 2e^{6t}q(8+6q+q^2) \nonumber \\
&\quad+ 3e^{2(2+q)t}(-8-6q+3q^2+q^3) \Big], \\
C_{\tilde{\mathcal{X}}}(t) &= \frac{e^{-2(2+q)t}}{24} \left( -1 + e^{2t} \right)^2 \left( 1 + 2e^{2t} + 2e^{6t} + e^{8t} \right), \\
C_{\tilde{\mathcal{S}}'} (t)&= \frac{e^{-2(2+q)t}}{3(-2+q)(-1+q)q(2+q)(4+q)} \Big[ q(2-3q+q^2) + 12e^{4t+qt}(-2+q+q^2)- 3e^{2(2+q)t}(-8+2q+q^2)  \nonumber \\
&\quad - e^{6t}q(8+6q+q^2) \Big], \\
C_{\mathcal{T}}(t) &= \frac{e^{-2(2+q)t}}{12(-2+q)q(2+q)(4+q)} \Big[ - q(-4+q^2) - 6e^{6t+qt}q^2(4+q)  - 6e^{4t}(-8+2q+q^2) - 2e^{6t}q(8+6q+q^2) \nonumber \\
&\quad+ 3e^{8t}q(8+6q+q^2) + 6e^{4t+qt}(-8+4q^2+q^3) \Big], \\
C_{\mathcal{S}'}(t) &= \frac{e^{-2(2+q)t}}{12(-2+q)q(4+q)} \Big[ 24e^{4t+qt} - (-2+q)q - 2e^{6t}q(4+q) + 3e^{4t}(-8+2q+q^2) \Big], \\
C_{\mathcal{T}'}(t) &= \frac{e^{-2(2+q)t}}{24(-2+q)(2+q)(4+q)} \Big[ 8 - 2q^2 - 12e^{4t+qt}(2+q) - 12e^{6t+qt}(4+q) - 3e^{4t}(-8+2q+q^2) \nonumber \\
&\quad + 2e^{6t}(8+6q+q^2) + 3e^{8t}(8+6q+q^2) \Big], \\
C_{\mathcal{S}''} (t)&= \frac{e^{-2(2+q)t}}{24(-2+q)(2+q)(4+q)} \Big[ 8 - 2q^2 - 12e^{4t+qt}(2+q)  + 12e^{6t+qt}(4+q) + 3e^{4t}(-8+2q+q^2) \nonumber \\
&\quad + 2e^{6t}(8+6q+q^2) - 3e^{8t}(8+6q+q^2) \Big], \\
C_{\mathcal{X}'} (t)&= \frac{e^{-2(2+q)t}}{24} \left( -1 + e^{6t} \right)^2, \\
C_{\mathcal{X}''}(t) &= -\frac{e^{-2(2+q)t}}{24} \left( -1 + e^{4t} \right)^3.
\end{align}
for the orthogonal case. The moment operator for the symplectic cases is obtained by the duality transformation $t \to -t$, $q \to -q$, ${\cal S}_4 \to -{\cal S}_4$, ${\cal X}_4 \to -{\cal X}_4$, ${\cal T}_4$ to $-{\cal T}_4$, ${\cal S}'_4 \to -{\cal S}'_4$, ${\cal T}'_4$ to $-{\cal T}'_4$, ${\cal S}''_4$ to  $-{\cal S}''_4$, and ${\cal X}''_4$ to $-{\cal X}''_4$. The trace moments for $k=4$ are
\begin{align}
\mathcal{R}_{1,1;1,1} =&\, \frac{3}{q^4} + \frac{1}{24q^4} e^{-2(\pm 2+q)t} \Big[ 72 e^{(q\pm 6)t} (\mp 1+q)q + 72 e^{(q\pm 4)t} (-2\pm q+q^2) + 4 e^{\pm 6t} q(\mp 6-7q+q^3) \nonumber\\
& +  e^{\pm 12t} q(\mp 6+11q\mp 6q^2+q^3)  + 9 e^{\pm 8t} q(\pm 6-5q\mp 2q^2+q^3) + q(\mp 6-q\pm 6q^2+q^3)\nonumber \\
& + 9 e^{\pm 4t} (8\mp 2q-9q^2\pm 2q^3+q^4) \Big], \nonumber\\
\mathcal{R}_{1,1;2} =&\, \pm \frac{1}{q^3} \mp \frac{1}{24q^3} e^{-2(\pm 2+q)t} (-1\pm q) \Big[ -24 e^{(q\pm 4)t} (2\pm q) + e^{\pm 12t} q(\pm 6-5q\pm q^2)  + 3 e^{\pm 8t} q(\mp 6-q\pm q^2) \nonumber\\
&- q(\pm 6+7q\pm q^2) - 3 e^{\pm 4t} (-8\mp 6q+3q^2\pm q^3) \Big], \nonumber\\
\mathcal{R}_{2,2} =&\, \frac{3}{q^2} + \frac{1}{24q^2} e^{-2(\pm 2+q)t} \Big[ -24 e^{(q\pm 6)t} (\mp 1+q)q + 24 e^{(q\pm 4)t} (-2\pm q+q^2) + 4 e^{\pm 6t} q(\mp 6-7q+q^3) \nonumber\\
&+ e^{\pm 12t} q(\mp 6+11q\mp 6q^2+q^3)  - 3 e^{\pm 8t} q(\pm 6-5q\mp 2q^2+q^3)  - 3 e^{\pm 4t} (8\mp 2q-9q^2\pm 2q^3+q^4) \Big] \nonumber\\
&+ q(\mp 6-q\pm 6q^2+q^3), \nonumber\\
\mathcal{R}_{1,3} =&\, \pm \frac{1}{24q} e^{-2(\pm 2+q)t}  \Big[ -6\mp q+6q^2\pm q^3 - 2 e^{\pm 6t} (-6\mp 7q\pm q^3) + e^{\pm 12t} (-6\pm 11q-6q^2\pm q^3) \Big], \nonumber\\
\mathcal{R}_{4} =&\, \pm \frac{1}{q} \mp \frac{1}{24q} e^{-2(\pm 2+q)t} (-1\pm q) \Big[ e^{\pm 12t} q(\pm 6-5q\pm q^2) - 3 e^{\pm 8t} q(\mp 6-q\pm q^2) \nonumber\\
&- q(\pm 6+7q\pm q^2) + 3 e^{\pm 4t} (-8\mp 6q+3q^2\pm q^3) \Big],
\end{align}
where the upper sign refers to the orthogonal case and the lower sign to the symplectic case.
\end{widetext}

\section{Moments of $V_k$}
\label{app:2}

To calculate the moment operators of Eq.\ (\ref{eq:V1}) we expressed the moment operators in terms of the trace moments, which are easy to calculate for the matrix $V$ of Eq.\ (\ref{eq:V}), because they depend on the eigenvalues of $V$ only. In the same way, we find an explicit expression for ${\cal V}_k$ for $k=4$,
\begin{align}
  \mathcal{V}_4=&\frac{1}{q(q+2)(q+4)(q+6)}
    \nonumber \\ &\, \mbox{} \times [(q^4+4q^3-28q^2-96q+16)\mathcal{I}_4
    \nonumber \\ &\, \mbox{}
    +4(q^2+6q+4)(\mathcal{S}_4+\mathcal{X}_4)
    -8(q+4)(\mathcal{T}'_4+\mathcal{X}'_4)
    \nonumber \\ &\, \mbox{}
    +16(\mathcal{\widetilde S}_4+\mathcal{\widetilde X}_4+\mathcal{\widetilde S}_4+\mathcal{T}_4+\mathcal{S}'_4+\mathcal{S}''_4+\mathcal{X}''_4)].
\end{align}
The tensors ${\cal I}_4$, ${\cal S}_4$, ${\cal X}_4$, $\tilde {\cal S}_4$, $\tilde {\cal X}_4$, $\tilde {\cal S}'_4$, ${\cal T}_4$, ${\cal S}'_4$, ${\cal T}'_4$, ${\cal S}''_4$, ${\cal X}'_4$, and ${\cal X}''_4$ are defined in App.\ \ref{app:1}.

We now describe a framework to calculate ${\cal V}_k$ for arbitrary $k$.
Hereto, we first note that the probability distribution of the ensemble $V$ of Eq.\ (\ref{eq:V}) is invariant whether $O$ is drawn from the special orthogonal group $\mbox{SO}(q)$ or the full orthogonal group $\mbox{O}(q)$. (To see this, note that right-multiplying $O$ by the $\text{diag}(-1, 1, \dots, 1)$ maps between the two disconnected components of ${\rm O}(q)$ without changing $V$.) We can explicitly rewrite $V$ as a Householder reflection,
\begin{equation}
  V_{ij} = \delta_{ij} - 2u_i u_j,
\end{equation}
where the vector $u_i = O_{i1}$ is the first column of the orthogonal matrix $O$. Because $O$ is Haar-distributed, $u$ acts as a uniformly distributed unit vector on the hypersphere $S^{q-1}$. To evaluate the $k$-th moments of $V$, we rely on the standard statistical moments of a uniform spherical vector. The expectation of a product of the components of $u$ is given by a sum over perfect matchings:
\begin{equation}
  \left\langle \prod_{m=1}^{2p} u_{i_m} \right\rangle = c_p \sum_{\pi \in \mathcal{M}_{2p}} \Delta_\pi(\mathbf{i}) ,   
\end{equation}
where $\mathcal{M}_{2p}$ is the set of perfect matchings of $2p$ indices, $\Delta_\pi(\mathbf{i})$ is the product of Kronecker deltas pairing the indices according to the matching $\pi$, and the spherical normalization constant is $c_p = [q(q+2)\cdots(q+2p-2)]^{-1}$. Because the perfect matchings $\mathcal{M}_{2p}$ form the basis of the Brauer algebra $\mathcal{B}_k$, the expectation value of the tensor power $V^{\otimes k}$ can be expanded entirely in terms of Brauer diagrams. Let $D \in \mathcal{B}_k$ represent a specific Brauer diagram, and let $l(D)$ denote the number of horizontal arcs within that diagram. Substituting the spherical moments into the binomial expansion of $(I - 2uu^{\rm T})^{\otimes k}$ allows us to state the exact expected tensor product as:
\begin{equation}
  \langle V^{\otimes k} \rangle = \sum_{D \in \mathcal{B}_k} \mathcal{A}_{k, l(D)} D,  
\end{equation}
where the combinatorial coefficient $\mathcal{A}_{k,l}$ depends solely on the total degree $k$ and the number of horizontal lines $l$, given by:
\begin{equation}
    \mathcal{A}_{k,l} = \sum_{r=0}^{l} \binom{l}{r} (-2)^{k-l+r} c_{k-l+r}.
\end{equation}
\bibliography{brownian}
\end{document}